\documentclass[aps,
pra,
twocolumn,
reprint,
noeprint,
superscriptaddress,
]
{revtex4-2}
\usepackage{amsmath}
\usepackage{amssymb}
\usepackage{CJK}
\usepackage{graphicx}% Include figure files
\usepackage[colorlinks,linkcolor=blue, anchorcolor=blue,citecolor=blue]{hyperref}
\usepackage{color}
\usepackage{xcolor}
\usepackage[toc,page,title,titletoc,header]{appendix}
\usepackage{dcolumn}% Align table columns on decimal point
\usepackage{bm}% bold math
\usepackage{float}
\usepackage{booktabs}
\definecolor{antiquefuchsia}{rgb}{0.57, 0.36, 0.51}
\definecolor{wrongultramarine}{rgb}{0.07, 0.04, 0.56}

\newcommand{\blue}[1]{\textcolor[rgb]{0.00,0.00,1.00}{#1}}

\newcommand{\bnabla}{\boldsymbol{\nabla}}
\newcommand{\Rb}[1]{${}^{#1}$Rb}
\newcommand{\Na}[1]{${}^{#1}$Na}

\newcommand{\ele}[2]{$^{#2}\rm{#1}$}
\newcommand{\cev}[1]{\reflectbox{\ensuremath{\vec{\reflectbox{\ensuremath{#1}}}}}}
\begin{document}
\title{Miscibility of Binary Bose-Einstein Condensates with $p$-wave Interaction}

\author{Min Deng}
\thanks{These authors contributed equally to this work.}
% \thanks{These authors contributed equally to this work.\\Corresponding author. \href{mailto:mxue@nuaa.edu.cn}{mxue@nuaa.edu.cn}}
\affiliation{College of Physics, Nanjing University of Aeronautics and Astronautics, and Key Laboratory of Aerospace Information Materials and Physics (NUAA), MIIT, Nanjing 211106, China}
\affiliation{College of Advanced Interdisciplinary Studies and Interdisciplinary Center of Quantum Information, National University of Defense Technology, Changsha 410073, China}

\author{Ming Xue {\small\textsuperscript{\blue{*}}}}
\email[Corresponding author. ]{mxue@nuaa.edu.cn}
% \thanks{These authors contributed equally to this work.\\Corresponding author. \href{mailto:mxue@nuaa.edu.cn}{mxue@nuaa.edu.cn}}
\affiliation{College of Physics, Nanjing University of Aeronautics and Astronautics, and Key Laboratory of Aerospace Information Materials and Physics (NUAA), MIIT, Nanjing 211106, China}

\author{Jinghan Pang}
\affiliation{College of Physics, Nanjing University of Aeronautics and Astronautics, and Key Laboratory of Aerospace Information Materials and Physics (NUAA), MIIT, Nanjing 211106, China}

\author{Hui Luo}
\affiliation{College of Advanced Interdisciplinary Studies and Interdisciplinary Center of Quantum Information, National University of Defense Technology, Changsha 410073, China}

\author{Zhiguo Wang}
\affiliation{College of Advanced Interdisciplinary Studies and Interdisciplinary Center of Quantum Information, National University of Defense Technology, Changsha 410073, China}

\author{Jinbin Li}
\affiliation{College of Physics, Nanjing University of Aeronautics and Astronautics, and Key Laboratory of Aerospace Information Materials and Physics (NUAA), MIIT, Nanjing 211106, China}

\author{Dayou Yang}
\affiliation{Institut f{\"u}r Theoretische Physik and IQST, Universit{\"a}t Ulm, Albert-Einstein-Allee 11, D-89069 Ulm, Germany}

\date{\today}

\begin{abstract}
We investigate the ground-state phase diagram of a binary mixture of Bose-Einstein condensates (BECs) with competing interspecies $s$- and $p$-wave interactions.
Exploiting a pseudopotential model for the $l=1$ partial wave, we derive an extended Gross-Pitaevskii (GP) equation for the BEC mixture that incorporates both $s$- and $p$-wave interactions. Based on it, we study the miscible-immiscible transition of a binary BEC mixture in the presence of interspecies $p$-wave interaction, by combining numerical solution of the GP equation and Gaussian variational analysis. Our study uncovers a dual effect---either enhance or reduce miscibility---of positive interspecies $p$-wave interaction, which can be precisely controlled by adjusting relevant experimental parameters. By complete characterizing the miscibility phase diagram, we establish a promising avenue towards experimental control of the miscibility of binary BEC mixtures via high partial-wave interactions. 
\end{abstract}

\maketitle
% \end{CJK*}

\section{Introduction}
Quantum gas mixtures stand as a versatile platform for ultracold collision research, quantum simulation and quantum information processing across diverse atomic species~\cite{PhysRevA.102.043301,PhysRevLett.103.140401,PhysRevLett.122.105302,mestrom21threebody,elliott2023quantum,Sheng22prl, Modugno02TwoAtomic,Hadzibabic02Two-Species,Thalhammer08DoubleSpecies,PhysRevResearch.1.033155,Sowinski_2019}. A paramount prerequisite for these pursuits is the precise control of the miscibility of cold gases, which is intricately influenced by the interplay between atomic interactions and experimental geometry~\cite{pu1998properties,li19spontan,lahaye2008d,Ravisankar_2021,2022Shell,PhysRevResearch.3.033247}. 

The miscibility of a binary Bose-Einstein condensate (BEC) mixture with $s$-wave interactions have been extensively
studied~\cite{Timmermans1998Phase,stenger1998spin,Esry1999Spontaneous,kim2002characteristic,Thalhammer08DoubleSpecies,tojo2010controlling}. The mixtures can be in either a miscible (M) or an immiscible (IM) phase,
dependent on the strengths of inter- and intra-species interactions, particle number ratio, trap potential and geometry~\cite{wen2012controlling,2020Phase,wen21pra,lee21miscibility}.
In the M phase, a homogeneous solution forms, while the IM phase exhibits a ball-and-shell or side-by-side condensate structure~\cite{myatt97production,hall1998dynamics,2008Hartree}. Under the Thomas-Fermi approximation (TFA), the transition is determined by $(a^c_{12})^2= a_{11}a_{22}$, where $a_{11(22)}$ is intraspecies $s$-wave scattering lengths of condensate 1(2), and $a^c_{12}$ is the critical interspecies s-wave scattering length for phase separation. Manipulating this transition in binary BECs is achievable via Feshbach resonance (FR), enabling control of $s$-wave interaction strengths~\cite{Chin2010Feshbach,Wu2012Optical}.

Exploiting beyond-$s$-wave broad FRs to enhance high partial-wave interactions, making them competitive with $s$-wave interaction thus creating novel states of matter,
is of significant experimental interest~\cite{2007p,yao2019degenerate,2016Feshbach,2017Controlled,Wang_2021}.
A comprehensive theoretical study of broad $s$-, $p$-, and $d$-wave FRs in various combinations of stable alkali-metal atoms was presented in Ref.~\cite{Cui2018Broad}.
Using Feshbach loss spectroscopy, new broad $s$-, $p$-, and $d$-wave FRs have been identified in diverse atomic mixtures, including Bose-Bose (\ele{K}{41}-\ele{K}{41}, \Na{23}-\Rb{87} and \Rb{85}-\Rb{87})~\cite{wang2015double,guo2021tunable,Dong2016Observation,Cui2017Observation,liu2018feshbach,yao2019degenerate}, 
Fermi-Fermi (\ele{Li}{6}-\ele{K}{40})~\cite{Chin2010Feshbach,Regal2004Observation,wille2008exploring,tiecke2010broad},
and Bose-Fermi (\ele{K}{41}-\ele{Li}{6}, \Na{23}-\ele{K}{40}) \cite{liu2018feshbach,zhu2017feshbach}.
Notably, the \Na{23}-\ele{K}{40} mixture exhibits a broad $d$-wave FR near 283 Gauss accompanied by a slightly narrower $s$-wave resonance~\cite{zhu2017feshbach}, and the \Rb{85}-\Rb{87} mixture features a broad $p$-wave FR with an associated $s$-wave resonance near 260 Gauss~\cite{Dong2016Observation}. The coexistence of different partial wave FRs motivates the exploration of experimental strategies that leverage multiple resonances to control both inter- and intra-species interactions~\cite{d2009ultracold,rui2017controlled,knoop2010magnetically}.

In this manuscript, we investigate the miscibility of a binary mixture of BECs with competing $s$- and $p$-wave interspecies interactions, 
shedding light on how these interactions can shape the ground state miscibility phase diagram. 
Previous studies extensively explored $p$-wave interactions in fermionic systems or Bose-Fermi mixtures, 
while studies on $p$-wave interactions in degenerate bosonic gases have been limited to 
heteronuclear mixtures due to restrictions imposed by quantum statistics.
In contrast to the repulsive $s$-wave interspecies interaction ($a_{12}>0$) which always results in energy penalty 
to the M phase and favors the IM phase, positive $p$-wave interspecies interaction has more intricate effects. 
Dependent on the specific overlapping configuration of the two condensates, 
the $p$-wave interaction can contribute either positive or negative mean-field energy to the mixture, 
thus favor either separation or mixing of the condensates. 
Such complexity introduces novel characteristics to BEC miscibility compared to the conventional $s$-wave scenario, 
offering the intriguing possibility of manipulating miscibility through higher partial-wave interactions. 
We investigate these novel aspects via numerical solution of the Gross-Pitaevskii equation (GPE), 
incorporating the $p$-wave interaction, and consolidate it with Gaussian variational analysis. 
As a concrete application of our theory, we study the miscibility phase diagram of an optically trapped 
ultracold mixture of \Rb{87}-\Na{23} atoms near a relatively broad $p$-wave FR around 284 Gauss~\cite{wang2013observation}. 
Such a condensate mixture has a long lifetime due to its positive intra- and inter-species $s$-wave scattering lengths~\cite{volz2003characterization,knoop2011feshbach}. 
We assume these background $s$-wave interactions remain constant in the vicinity of the $p$-wave resonance under consideration, 
a reasonable approximation for the broad resonance and dilute limits considered here.

The rest of the manuscript is structured as follows. Section~\ref{sec:model} outlines the theoretical model, starting with the derivation of the  $p$-wave interaction model for a binary BEC mixture in Sec.~\ref{sec:pwavemodel}, followed by the establishment of an extended mean-field GP equation in Sec.~\ref{sec:pwavemodelGP}. To gain insight into the ground-state phases, we employ a simple Gaussian variational ansatz in Sec.~\ref{sec:NumericalmodeloneDim}
for one-dimensional (1D) condensates. 
In Sec.~\ref{sec:ThegroundstatePD}, we present the phase diagrams for two-dimensional (2D) and three-dimensional (3D) regimes. Lastly, in Sec.~\ref{sec:conclusion} we conclude with discussions.

\section{THEORETICAL MODEL}\label{sec:model}
\subsection{$p$-wave interaction model}\label{sec:pwavemodel}
Modeling two-body interactions constitutes a fundamental step in developing theories for many-body systems.
In the ultracold regime, atomic collisions dominated by $s$-wave scattering,
which can be modeled accurately with a Fermi-Huang pseudopotential~\cite{huang1957quantum,1987Statistical}.
The presence of non-$s$-wave scattering resonances transforms this scenario and requires tailoring pseudopotentials to these higher partial waves. Here, we employ a concise single-channel $p$-wave pseudopotential model derived by Idziaszek and Calarco~\cite{idziaszek2006pseudopotential,Zbigniew09analytical},
\begin{eqnarray}\label{eq:pseudopotential}
	V_{p}(\mathbf{r}) =  \frac{\pi\hbar^2 a_p^3}{m_r}\cev{\bnabla}\delta({\mathbf r})\cdot\frac{\partial^3}{\partial r^3}r^3\vec\bnabla,
\end{eqnarray}
where $\cev{\bnabla}$ ($\vec{\bnabla}$) denotes the gradient operator acting on the left (right) of the pseudopotential,
$a_p$ is the energy-dependent scattering length for $p$-wave interaction, satisfying $a_p^3(k)=-\tan\delta_1(k)/k^3$ with $\delta_1$ the $p$-wave scattering phase shift, and $k$ is the wave vector associated with the collision energy $E$. Such a single-channel pseudopotential model is applicable if the splitting of the $p$-wave FR peak due to magnetic dipole interactions is small such that the angular anisotropy of the interaction can be neglected. It has been successfully applied, e.g., in confirming geometric resonances~\cite{granger2004tuning,olshanii1998atomic}, fermion scattering within quasi-2D systems~\cite{idziaszek2006pseudopotential} and investigating trap-induced shape resonances in $p$-wave interactions of ultracold atoms~\cite{reichenbach2006quasi}.

Denoting the atomic fields of the condensate mixture as $\hat{\boldsymbol{\psi}}(\mathbf{r})=[\hat{\psi}_1(\mathbf{r}), \hat{\psi}_2(\mathbf{r})]^T$. 
The second-quantized form of the pseudopotential Eq.~\eqref{eq:pseudopotential} can be written as
\begin{equation}
\hat{V}_p = \frac{\pi\hbar^2 \nu_p}{m_r}\!\!\!\int\!\! d^3r
(\hat{\psi}_2^\dagger\nabla\hat{\psi}^\dagger_1-\hat{\psi}_1^\dagger\nabla\hat{\psi}^\dagger_2)
(\hat{\psi}_2\nabla\hat{\psi}_1-\hat{\psi}_1\nabla\hat{\psi}_2),\label{eq:Vp}
\end{equation}
where $\nu_p = a_p^3(k)$ is the $p$-wave scattering volume and $m_r$ denotes the reduced mass of the two atom species. 
We note that Eq.\,(\ref{eq:Vp}) can also be derived based on $p$-wave interaction model that involves $p$-wave molecular degrees of freedom (DOFs) via adiabatic elimination of the molecular fields~\cite{li19spontan}.

\subsection{Extended Gross-Pitaevskii equation with interspecies $p$-wave interaction}\label{sec:pwavemodelGP}
To describe the ground state property of the BEC mixture, we use a mean-field Gross-Pitaevskii equation that incorporates the interspecies $p$-wave interaction. The mean-field energy functional governing a 3D BEC mixture is given by~\cite{leo09pwave}:
\begin{eqnarray}\label{eq:eqEnergy}
E\left[\boldsymbol{\psi}^*,\boldsymbol{\psi}\right]&=&\int d^3{r}\sum^2_{i=1}\left(\frac{\hbar^2}{2m_i}|\nabla\psi_i|^2+V_T^i|\psi_i|^2\right)\nonumber\\
&&+\int\! d^3{r}\left(\sum_{i=1}^2\frac{g_i}{2}|\psi_i|^4+g_{12}|\psi_1|^2|\psi_2|^2\right)\nonumber\\
&&+\int\! d^3{r}\frac{\pi\hbar^2\nu_p}{m_r}|\psi_2\nabla\psi_1-\psi_1\nabla\psi_2|^2,\label{eq:variantion}
\end{eqnarray}
where the wavefunction is normalized as $\int d^3r|\psi_i|^2=N_i$ for the two species $i=1,2$.
The ground-state wavefunctions can be obtained via the variational extremum condition
\begin{eqnarray}
	{\delta}\left[E-\mu_iN_i\right]/\delta\psi_i^*=0,\label{eq:deltaE}
\end{eqnarray}
where $\mu_i$ is the chemical potential of species $i$. The first line of Eq.~(\ref{eq:variantion}) describes the kinetic energy and harmonic trapping potential of the two species, with $V_T^i=m_i(\omega_{ix}^2x^2 + \omega_{iy}^2y^2+ \omega_{iz}^2z^2)/2$, and $m_i$ and $\omega_i$ being the mass and trapping frequency of the species $i$. In the second line, 
$g_i=4\pi\hbar^2a_i/m_i$ denotes the $s$-wave intraspecies interaction strength, and $g_{12}=2\pi\hbar^2a_{12}/m_r$ represents the interspecies interaction strength,
where $a_i(a_{12})$ denotes the $s$-wave intra-(inter-)species scattering length, and $m_r=m_1m_2/(m_1+m_2)$ is the reduced mass.

Introducing the characteristic length scale $l_0=\sqrt{\hbar/m_1\omega_{1x}}$
and energy scale $\varepsilon_0= \hbar\omega_{1x}$,
and redefining the normalized mean-field wave function $\phi_i=l_0^{3/2}N_i^{-1/2}\psi_i$ ($i=1,2$),
the GP equations derived from Eqs.~(\ref{eq:eqEnergy}) and (\ref{eq:deltaE}) read
\begin{eqnarray}
\mu_1\phi_1=
&&\left[-\frac{\nabla^2}{2}+\sum_{\sigma}\frac{\xi_{1\sigma}^2\sigma^2}{2}
+\beta_1|\phi_1|^2+\beta_{12}|\phi_2|^2\right]\phi_1\nonumber\\
&&-\beta_p|\phi_2\nabla\phi_1-\phi_1\nabla\phi_2|^2,\label{eq:eqEnergy3}\\
\mu_2\phi_2=
&&\left[-\frac{\alpha_m\nabla^2}{2}+\sum_{\sigma}\frac{\xi_{2\sigma}^2\sigma^2}{2\alpha_m}
+\beta_2|\phi_2|^2+\gamma\beta_{12}|\phi_1|^2\right]\phi_2\nonumber\\
&&-\gamma\beta_{p}|\phi_1\nabla\phi_2-\phi_2\nabla\phi_1|^2,\label{eq:eqEnergy4}
\end{eqnarray}
where $\alpha_m=m_1/m_2$, $\gamma=N_1/N_2$, and
$\xi_{i\sigma}=\omega_{i\sigma}/\omega_{1x}$ with $\sigma=x,y,z$.
The parameters $\beta_i=4\pi a_i\alpha_m N_i/l_0$ and $\beta_{12}=2\pi \left(1+\alpha_m\right)a_{12} N_2/l_0$ are rescaled $s$-wave interaction strengths, and $\beta_p=\pi\left(1+\alpha_m\right) \nu_p N_2/l_0^{3}$ is the rescaled $p$-wave interaction strength.

Defining the dimensionless energy functional $\mathcal{E}={E}/{(N_1\varepsilon_0)}$, Eq.~(\ref{eq:eqEnergy}) can be recast as
\begin{eqnarray}\label{eq:eqEnergy2}
\mathcal{E}\!&=&\!\int\!{d^3}r'\!\!\sum_{j=1,2}\!\!\gamma^{1-j}[{\frac{1}{2}\alpha _m^{j - 1}{{|\nabla'{\phi _j}|}^2} + \alpha_m^{1 - j}{V_j}(r'){{|{\phi_j}|}^2}}]\nonumber\\
&& + \int\!{d^3}r'[\frac{{{\beta _1}}}{2}{\left| {{\phi _1}} \right|^4} + \frac{{{\beta _2}}}{{2\gamma }}{\left| {{\phi _2}} \right|^4} + {\beta _{12}}{\left| {{\phi _1}} \right|^2}{\left| {{\phi _2}} \right|^2}]\nonumber\\
 && + {\beta_p}\int{d^3}r'\,{{|{\phi_2}\nabla {\phi_1} - {\phi_1}\nabla {\phi_2}|}^2},
\end{eqnarray}
with the rescaled potential $V_j(r)=\sum_{\sigma=x,y,z}\xi_{j\sigma}^2 \sigma^2/2$.

To explore the interplay between $s$- and $p$-wave interactions, we employ the imaginary-time evolution method ~\cite{Dalfovo1996Bosons} to analyze the ground-state phase diagram of the BEC mixture. Notably, this method is not a global optimization strategy; it converges to local energy minima dependent on the initial trial wavefunction. Successful application necessitates an initial trial wavefunction encompassing all possible symmetries to ensure convergence to the genuine ground state. This consideration is crucial in our study, given that $p$-wave interactions may introduce excess metastable states (see Sec.~\ref{sec:VariationalModel}). We terminate the evolution upon reaching energy convergence criteria, specifically $\left| {\mathcal{E}\left({\tau_{n+1}}\right) - \mathcal{E}({\tau _{n}})} \right|/{{\mathcal{E}(\tau_{n})}}<{10^{-7}}$, where $\tau_n$ denotes discretized imaginary time. This criteria proves sufficient for attaining the genuine ground state in our investigation below.

\subsection[short]{Order parameters}
We adopted the following pair of order parameters to identify distinct ground-state configurations,
as introduced in previous studies~\cite{Lee2016Phase,Fava2018Observation,wen21pra}:
\begin{eqnarray}
\eta &=& \int d^3r\,|\phi_1||\phi_2|,\\
  d  &=& \left|\int d^3r\,\left(|\phi_1|^2-|\phi_2|^2\right)\mathbf{r}\right|.
\end{eqnarray}
Here, $\eta$ represents the density overlap between two BEC species, 
and the wave functions of the two species ($\phi_{i=1,2}$) are normalized to unit. 
Thus, the value is close to 1 when the two BECs overlap almost completely, and less than 1 otherwise, 
distinguishing between the miscible (M) and immiscible (IM) phases.
The parameter $d$ measures the center of mass displacement between the two condensates. 
Hence, the ground state resides in a symmetric immiscible (SIM) phase for $(\eta < 1, d = 0)$, 
typically exhibiting a ball-and-shell structure where one species forms a shell around the other. 
In contrast, the ground state resides in an asymmetric immiscible (AIM) phase for $(\eta < 1, d > 0)$, 
where the two condensates coexist side by side.

Before numerical calculations, let us qualitatively analyze the role of each term in Eq.~(\ref{eq:eqEnergy2}): 
(i) Both kinetic and potential energy favors the mixing of the two BECs, and they dominate the miscibility of the mixture for small $s$- and $p$-wave interactions.
(ii) The intraspecies $s$-wave interaction energy monotonically decreases with the extension of the condensates' wavefunction, thereby favoring mixing.
(iii) The repulsive $s$-wave interspecies interaction between the BECs favors the IM phase~\cite{hall1998dynamics,pu1998properties,stenger1998spin,tojo2010controlling,kim2002characteristic}.
In the limit $a_{12}\gg a_{11},a_{22}$, the condensates cease to overlap, resulting in complete phase separation.
(iv) The $p$-wave interspecies interaction is intricate, depending not just on the wave function but also its gradient. In the M phase, similar density profiles result in a small interaction energy ${\left| {{\phi _2}\nabla {\phi _1} - {\phi _1}\nabla {\phi _2}} \right|^2}$ [cf. last line of Eq.~\eqref{eq:eqEnergy2}]. As condensates move apart in space, this term gradually increases. Conversely, with minimal spatial overlap between the condensates, this term approaches zero. Thus, for $\nu_p>0$, $p$-wave interaction tends to either mix or entirely separate BECs. Conversely, for $\nu_p<0$, partially mixed configurations lowers the $p$-wave interaction energy.

\section{Miscibility phase diagram of quasi-one-dimentional BEC mixture}\label{sec:NumericalmodeloneDim}
	
\subsection{Numerical phase diagram} \label{sec:Numericalmodel}
We investigate the miscibility phase diagram of binary BEC mixture in a quasi-1D geometry via imaginary time evolution. To illustrate the essential physics,  we make the simplification assumption of equal masses ($m_1=m_2=1$) and trapping frequencies ($\omega_1=\omega_2=1$) for the two condensates. We initialize the condensates in two Gaussian wavefunctions of the same standard deviation $\sigma$, whose center is displaced by a distance $\delta_r$. We make diverse choices of $(\sigma, \delta_r)$ to guarantee 
that imaginary time evolution converges to the genuine ground state. 

\begin{figure}[t!]
	\centering
	\includegraphics[width=1.0\columnwidth]{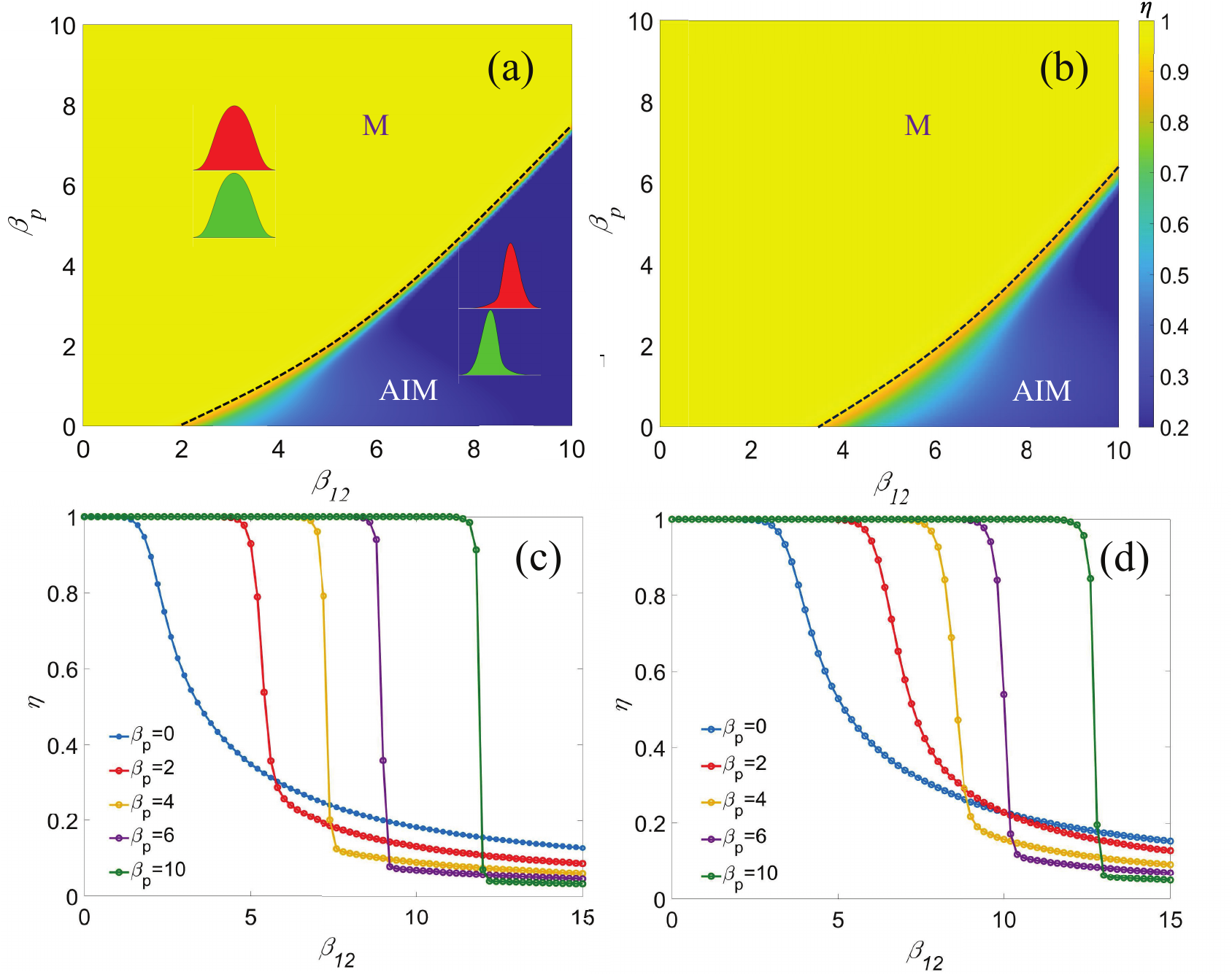}
	\caption{(a-b) Ground-state phase diagram of a binary mixture of BECs in 1D, as a function of the dimensionless interspecies $s$-wave interaction $\beta_{12}$ and $p$-wave interaction $\beta_p$. The diagram is calculated via numerical imaginary-time evolution of the GP equation. The schematic filled with red and green represent the density distributions of the BECs in the miscible (M) and immiscible (IM) phases. We consider in (a) $\beta_1 = \beta_2 = 0$, i.e., without intraspecies interactions; in (b) $\beta_1 = \beta_2 = 2$. Since the boundaries of order parameters $d$ and $\eta$ are identical, only the diagrams of $\eta$ are shown. (c-d) Dependence of the order parameter $\eta$ on the interspecies interaction $\beta_{12}$ for different  $\beta_p$. We consider in (c) $\beta_1 = \beta_2 = 0$ and in (d) $\beta_1 = \beta_2 = 2$.}%
	\label{fig1}%
\end{figure}

The resulting numerical phase diagram is shown in Fig.~\ref{fig1} against the effective $s$- and $p$-wave interspecies interactions $\beta_{12}$ and $\beta_p$. The mixture has two phases, the M phase and the AIM phase, with the black dashed line representing the phase boundary. In the AIM phase, the two BECs symmetrically separate in opposite directions, with $\eta<1$ and $d>0$. We show the order parameter $\eta$ only, as the phase boundary of $d$ perfectly aligns with that of $\eta$. Figs.~\ref{fig1}(a) and \ref{fig1}(b) show the ground-state phase diagram with and without $s$-wave intraspecies interactions ($\beta_{1}, \beta_{2}$) respectively, which are qualitatively similar.

When both $\beta_{12}$ and $\beta_p$ are small, i.e., 
$(\beta_{12},\beta_p)\lesssim (6.0, 4.0)$ in Fig.~\ref{fig1}(a)
and $(\beta_{12}, \beta_p) \lesssim (10.0, 6.0)$ in Fig.~\ref{fig1}(b), the order parameters
$\eta$ and $d$ various continuously  across the phase boundary, cf., the solid blue and red lines in Figs.~\ref{fig1}(c) and (d).
However, upon surpassing a threshold of the interaction strengths,
the order parameter exhibits discontinuity across the phase boundary, cf., the purple and green curves in Figs.~\ref{fig1}(c) and (d). This novel feature does not appear in the conventional $s$-wave miscibility phase diagram, thus is a direct result from the $p$-wave interspecies interaction. Its physical origin is investigated in the subsequent Sec.~\ref{sec:VariationalModel} via Gaussian variational analysis.

Comparing Figs.~\ref{fig1}(c) and (d), we observe that nonzero intraspecies interactions $\beta_{1,2}$ does not modify the qualitative features of the phase diagram---it only causes a slight expansion of the phase boundary towards the AIM phase region. For instance, when $\beta_p=0$, the threshold for $\eta$ to drop from 1 shifts from $\beta_{12} \approx 2.0$ to $\beta_{12}\approx 4.0$. This result aligns with the physical intuition that intraspecies interactions favor mixing of the condensates, necessitating a stronger $\beta_{12}$ for phase separation.

\subsection{Gaussian variational analysis} \label{sec:VariationalModel}
\begin{figure*}[t!]
\centering
\includegraphics[width= 2.0\columnwidth]{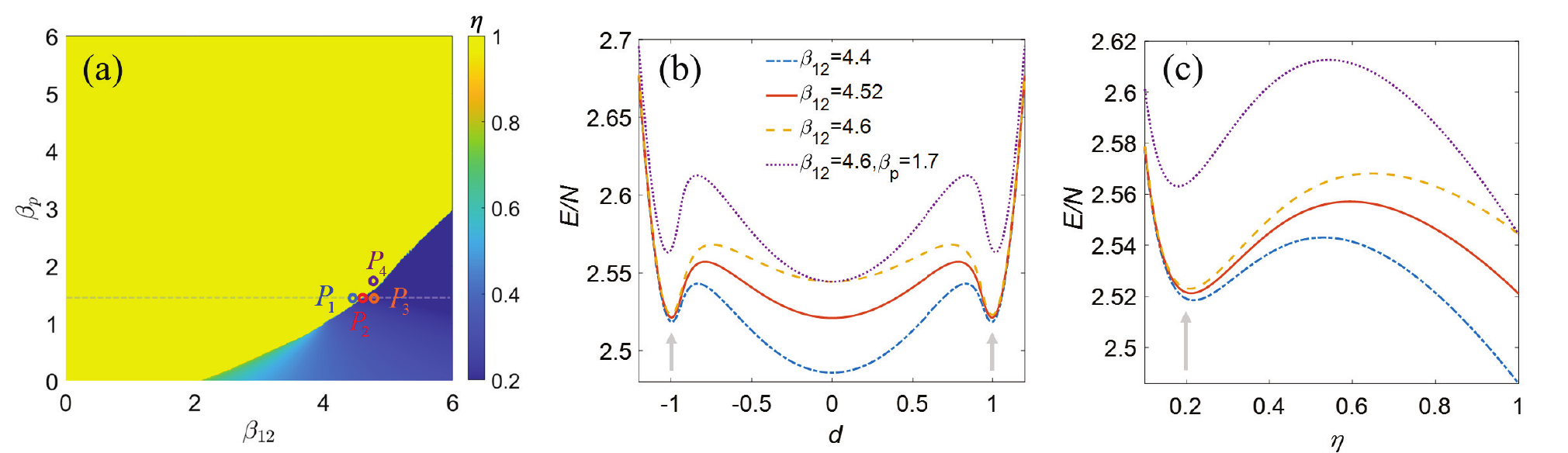}
\caption{(a) Ground-state phase diagram of a binary mixture of BECs in 1D, obtained via Gaussian variational analysis (cf. Sec.~\ref{sec:VariationalModel}). Several points $(\beta_{12},\beta{p})$ near the phase transition point $P_2=(4.52,1.5)$, including $P_1=(4.4,1.5)$, $P_3=(4.6,1.5)$, and $P_4=(4.6,1.7)$, are marked. These  points are used to illustrate the physical mechanism underlying the abrupt jumps of the order parameter across the phase boundary, cf. Sec.~\ref{sec:VariationalModel}. Dashed line corresponds to $\beta_p=1.5$. (b-c) Gaussian variational energy of the BEC mixture, $E/N$, at as a function of the order parameter $d$ [panel (b)] and $\eta$ [panel (c)], at points $P_1$ (blue dashed line), $P_3$ (orange solid line) and $P_4$ (purple dotted line). Light gray arrows indicate two possible metastable states with non-vanishing order parameter.}%
\label{fig2}%
\end{figure*}

We employ a Gaussian variational method to investigate further the ground-state phase diagram of the quasi-1D BEC mixture. This method is suitable for weak interactions and provides insights into single-component BECs and BEC mixtures~\cite{garcia96prl,perez1997dynamics,busch1997stability}. Without loss of generality, we set the intraspecies interaction ($\beta_{1,2}=0$), and assume equal masses ($m_1=m_2=1$) and trapping frequencies ($\omega_1=\omega_2=1$) of the two condensates as in Sec.~\ref{sec:Numericalmodel}. Under these assumptions, the numerical results in Sec.~\ref{sec:Numericalmodel} indicate that the two BECs displace symmetrically along opposite directions in IM phase. Hence, we define the following two Gaussian variational wavefunctions for the condensates,
\begin{eqnarray}
\phi_1(x)=\frac{1}{{\pi}^{1/4}{\sigma}^{1/2}}e^{{-\left( {x - {d}} \right)^2}/{2\sigma^2}},\\
\phi_2(x)=\frac{1}{{\pi}^{1/4}{\sigma}^{1/2}}e^{{-\left( {x + {d}} \right)^2}/{2\sigma^2}},
\end{eqnarray}
which suffice to capture the characteristics of both M and AIM phases. This allows us to express the order parameter $\eta=\exp\left(-d^2/{\sigma^2}\right)$ and the dimensionless energy functional in Eq.~(\ref{eq:eqEnergy2}) as
\begin{eqnarray}\label{eq:eqEnergyAnalysis}
\mathcal{E}/N&=&\frac{1}{2\sigma^2}+\frac{1}{2}(\sigma^2+2d^2)\nonumber\\
&&+\frac{\beta_{12}}{\sqrt{2\pi}}e^{-2d^2/\sigma^2}+\frac{4\beta_p d^2}{\sqrt{2\pi}\sigma^5}e^{-2d^2/\sigma^2}.
\end{eqnarray}
The ground-state phase diagram from numerical minimization of Eq.~\eqref{eq:eqEnergyAnalysis} is shown in Fig.~\ref{fig2}(a), which aligns excellently with Fig.~\ref{fig1}(a). Notably, when $\beta_p=0$, both results indicate an AIM phase entry threshold at $\beta_{12}\approx 2.0$.

Next, we analyze the discontinuity of the order parameter $\eta$ using the Gaussian variational approach.
Figs.~\ref{fig2}(b) and (c) display the behavior of average energy $\mathcal{E}/N$ in the vicinity of the phase boundary as a function of the order parameters $d$ and $\eta$. 
The discontinuity of $\eta$ across the phase boundary stems from the existence of two metastable states, as indicated by arrows in Figs.~\ref{fig2}(b) and (c) (with order parameters $d\approx\pm1.0$ and $\eta\approx0.2$). To further illustrate this, we consider two points $P_1=(4.4, 1.5)$ and $P_3=(4.6, 1.5)$, lying in the M and AIM phase respectively, in the vicinity of the phase transition point $P_2=(4.52,1.5)$, as shown in the phase diagram Fig.~\ref{fig2}(a). The blue dot-dashed line in Fig.~\ref{fig2}(b) corresponds to the point $P_1$ in the M phase, which indicates that the energy of the two metastable states is higher than the ground state (order parameter $d = 0, \eta=1$). However, the energy of the two metastable states decreases as we move from $P_1$ to $P_3$ by increasing $\beta_{12}$. When $\beta_{12}=4.6$, corresponding to point $P_3$ in Fig.~\ref{fig2}(a), 
the energy of the two metastable states becomes lowest,  as shown by the yellow-dashed line in Fig.~\ref{fig2} (b-c). Hence, the order parameters $d$ jumps abruptly from 0 to $\pm1$ and $\eta$ from 1 to 0.2 when crossing the transition point $P_2$.

Moreover, the $p$-wave interspecies interaction enhances mixing of the condensates. In Fig.~\ref{fig2}(b), as $\beta_p$ increases from 1.5 to 1.7,
the ground state undergoes a phase transition from the AIM phase at position $P_3$ (yellow-dashed curve) to the M phase at position $P_4$ (purple-dotted curve).
These two curves shows that the $p$-wave interaction introduces an energy penalty to the metastable states, thereby enhancing mixing. Consequently, to enter the AIM phase necessitates a larger $\beta_{12}$.

Next, we apply Landau's theory to elucidate the characteristics of the phase transition. To account for the metastable states, we construct a Landau free energy functional~\cite{landau2013statistical} by expanding the energy functional Eq.~(\ref{eq:eqEnergyAnalysis}) in power series of the order parameter $d$ up to the sixth order, assuming $\beta_p \leq 1$. We hence arrive at
\begin{equation} \label{eq:eqEnergyLandau}
	\varepsilon(d,\sigma) =\mathcal{E}/N= A + B{d^2} + C{d^4} + D{d^6},
\end{equation}
where we define
\begin{eqnarray}
A &=& \frac{1}{{2{\sigma ^2}}} + \frac{1}{2}{\sigma ^2} + \frac{{{\beta _{12}}}}{{\sqrt {2\pi } \sigma }},\\
B &=& 1 - \frac{{2{\beta _{12}}}}{{\sqrt {2\pi } {\sigma ^3}}} + \frac{{4{\beta _p}}}{{\sqrt {2\pi } {\sigma ^5}}},\\
C &=& \frac{2}{{\sqrt {2\pi } {\sigma ^7}}}\left( {{\beta _{12}}{\sigma ^2} - 4{\beta _p}} \right),\\
D &=& \frac{4}{{\sqrt {2\pi } {\sigma ^9}}}\left( {2{\beta _p} - \frac{1}{3}{\beta _{12}}{\sigma ^2}} \right).
\end{eqnarray}

We minimize the Landau free energy functional Eq.~(\ref{eq:eqEnergyLandau}) with respect to $d$ and $\sigma$, in order to identify the ground and metastable states of the mixture. We assume that near the phase transition, $\varepsilon(d,\sigma)$ gradually changes with $\sigma$ near the phase transition point, whereas $d$ can jump abruptly from zero to nonzero values in the regime $\eta<1$. As a result, $\sigma$ can be treated as a constant determined solely by $\beta_{12}$ and $\beta_{p}$,
that is, $\sigma \simeq \sigma (\beta_{12},\beta_p)$.
$\varepsilon(d,\sigma)$ can then be expressed as a standard Landau's free energy functional that depends on $d$.
Notably, the inclusion of terms $\propto d^6$ in $\varepsilon(d,\sigma)$ allows it to account for three metastable states (cf. Fig.~\ref{fig2}).
To ensure that the energy has a lower bound, the condition $D> 0$ must be satisfied.
Utilizing ${\partial \varepsilon}/{\partial d} = 0$ and ${{\partial ^2}\varepsilon}/{{\partial^2}d} > 0$, we find the metastable states of the mixture,
\begin{equation}
d_0 = 0,\quad
d_ \pm ^2 = \frac{{ - C + \sqrt {{C^2} - 3BD} }}{{3D}}.
\end{equation}
It is clear that the presence of the metastable states requires ${C^2} - 3BD > 0$ and $C < 0$, otherwise only one stable state with $d_0 = 0$ exists, which is the ground state of the mixture.

In the presence of metastable states, the phase transition is of first order. The system's ground state is the M phase ($d= d_0=0 $)
when $\varepsilon(d_0) < \varepsilon(d_\pm)$; otherwise, it is in AIM phase ($d = d_\pm $). Thus, the first-order phase transition point satisfies the equation $\varepsilon(d_0) = \varepsilon(d_\pm)$, 
which simplifies to $C^2 = 4BD$.

Summarizing the above analysis, we have approximately determined the phase transition characteristics of the binary BEC mixture. When $C<0$ and $D>0$, the ground state is in the M phase if the condition $3BD<C^2<4BD$ is met. In contrast, if $C^2>4BD$, the ground state is in the IM phase and the mixing configuration becomes metastable. The phase boundary corresponds to the equation $C^2 = 4BD$. By expressing parameters $B, C, D$ in terms of $\beta_{12}$, $\beta_{p}$, and $\sigma(\beta_{12}, \beta_{p})$, we can derive an approximate formula for the phase boundary. In the regime that $\beta_{12}$ and $\beta_{p}$ are small ($\sigma \simeq 1$), the phase boundary formula reduces to
\begin{equation}
5\beta_{12}^2 - 4(10\beta_p + \sqrt{2\pi})\beta_{12} + 24\beta_p(2\beta_p + \sqrt{2\pi}) = 0.
%\nonumber
\end{equation}
Notably, when $\beta_{p} = 0$, this equation yields $\beta_{12}=\frac{4}{5}\sqrt {2\pi}\approx 2$,
which closely matches the numerical results in Fig.~\ref{fig1}(a) and Fig.~\ref{fig2}(a).

\section{MISCIBILITY PHASE DIAGRAM in two- and three-dimension} \label{sec:ThegroundstatePD}

\subsection{Ground-state phase diagram without intraspecies interactions}\label{sec:ThegroundstatePDwithoutintra int}
\begin{figure*}[t!]
	\centering
	\includegraphics[width= 2.\columnwidth]{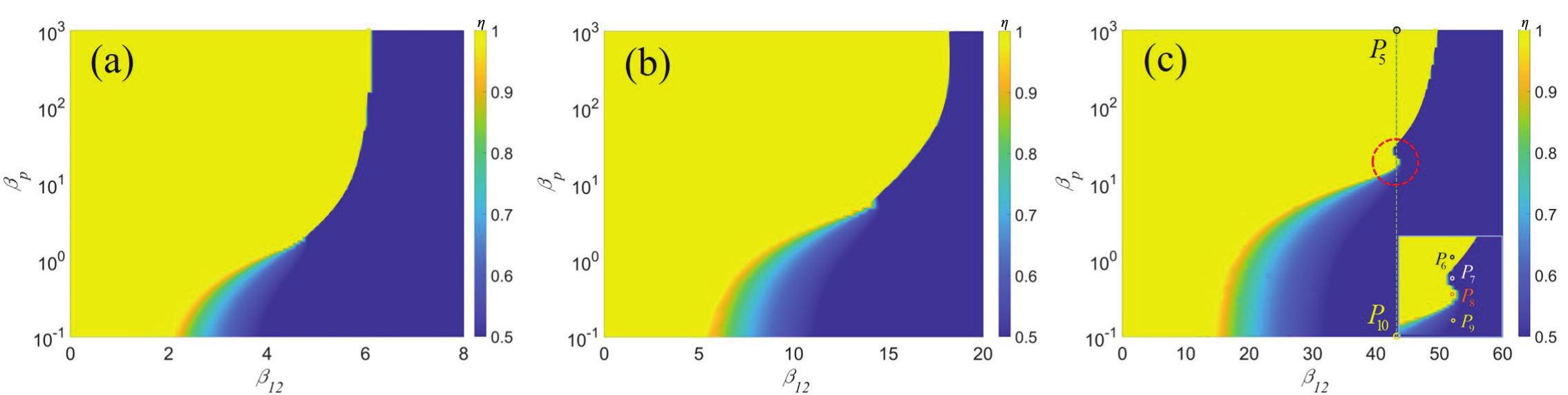}
	\caption{Ground-state phase diagram of a binary mixture of BECs in (a) 1D, (b) 2D and (c) 3D, as a function of the dimensionless interspecies s-wave interaction $\beta_{12}$ and $p$-wave interaction $\beta_p$. As the boundaries of order parameters $d$ and $\eta$ are identical, only the distribution of $\eta$ is ploted. In the pure yellow region, the system in a completely M phase ($\eta = 1$ and $d = 0$), while in other regions, it is in an AIM phase($\eta < 1$ and $d > 0$). In (c), along the line $\beta_{12} = 43$, several points from $P_5=(43, 10^3)$-$P_{10}=(43, 0)$ are marked. They offer insights into the dual effects of $p-$wave interaction for phase separation, as analyzed in Sec.~\ref{sec:ThegroundstatePDwithoutintra int}. Insert of panel (c) zooms in the region inside the red dashed circle of panel (c). }%
	\label{fig3}%
\end{figure*}
Extending our previous analysis, we explore below the ground-state phase diagram of a binary mixture of BECs in 2D and 3D geometry. 
In higher spatial dimensions, metastable states become more prevalent, 
posing challenges to the use of imaginary time propagation methods for obtaining the true ground state. 
To address this, we exploit two trial wavefunctions, one with spherical symmetry ($\delta_r=0$) and the other with asymmetry ($\delta_r=3\sigma$), encompassing all possible symmetry-broken configurations. Moreover, we temporarily exclude the $s$-wave intraspecies interactions to highlight the effect of interspecies interactions in this section.
We denote $^{87}$Rb as species 1 and $^{23}$Na as species 2, utilizing an isotropic harmonic trap.
For species 1 (Rb atoms), we set $\omega_1 = (2\pi)160$ Hz, while for species 2 (Na atoms),
we set $\omega_2=\omega_1\sqrt{m_1/m_2}\approx 2\omega_1$.
This choice ensures equal effective trap potentials despite the differing atomic masses~\cite{Esry1999Spontaneous}.

\begin{figure}[b!]
\centering
\includegraphics[width= 1\columnwidth]{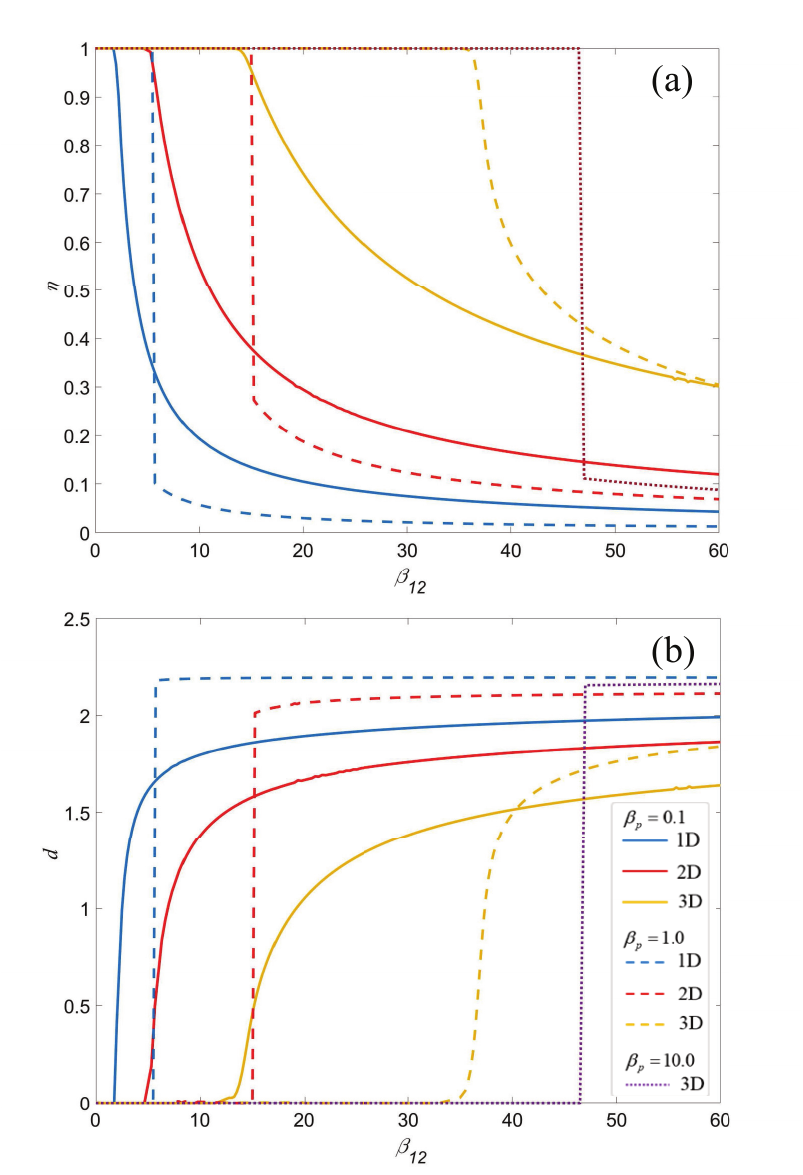}
\caption{Dependence of the order parameters $\eta$ (a) and $d$ (b) on the interspecies $s$-wave interaction strength $\beta_{12}$, for various interspecies $p$-wave interaction strength $\beta_{p}$ and spatial dimension. Solid line: $\beta_p=0.1$; dashed lines: $\beta_p=1$; purple dotted line: $\beta_p=10.0$ and in 3D. }%
\label{fig4}%
\end{figure}

\begin{table*}[t!]
\caption{Characterization of the ground-state phase and energy of the BEC mixture at location 5-10 along the vertical dashed line $\beta_{12}=43$ in Fig.~\ref{fig3}(c).
M(AIM) corresponds to the miscible(asymmetric immiscible) phase. $\varepsilon_{\rm tot}$ is the total energy of the mixture, $\varepsilon_{k_i}$ and $\varepsilon_{p_i}$ are the kinetic and potential energies of the component $i=1,2$. $\varepsilon_{\beta_{12}}$ and $\varepsilon_{\beta_p}$ are the $s$- and $p$-wave interspecies interaction energies. All energies are dimensionless.}
\label{Table}
\begin{tabular}{*{12}{c}}
	\toprule
	Location & Phase & \quad$\varepsilon_{\rm tot}$ &\quad $\eta$ &\quad $d$ & \quad$\varepsilon_{k_1}$ & \quad$\varepsilon_{k_2}$ & \quad$\varepsilon_{p_1}$  & \quad$\varepsilon_{p_2}$  & \quad$\varepsilon_{\beta_{12}}$ & \quad$\varepsilon_{\beta_p}$  &\quad Real Phase\qquad \\
	\midrule
	$P=(43, -)$ & M & 4.552 &\quad 1.000 &\quad 0.000 &\quad 0.464 & \quad 0.464& \quad 1.230 &\quad 1.230 & \quad 1.164 & \quad 0.000& \\
	\midrule
	$P_5=(43, 10^3)$ & AIM &\quad 4.720 &\quad 0.026  &\quad 2.226 &\quad 1.115 &\quad 1.115  & \quad 1.230 & \quad 1.230 & \quad 0.001&\quad 0.029& M \\
	$P_6=(43, 35)$ & AIM & \quad 4.564 &\quad 0.283  &\quad 1.997 &\quad 0.871 &\quad 0.871 &   \quad 1.241 &  \quad 1.241&\quad0.098 &\quad 0.242& M\\
	$P_7=(43, 27)$ & AIM & \quad 4.537 &\quad 0.356  &\quad 1.917 &\quad 0.805 &\quad 0.805  &   \quad 1.249 & \quad 1.249 &\quad0.153 &\quad 0.276& AIM\\
	$P_8=(43, 20)$ & AIM & \quad 4.578 &\quad 0.823  &\quad 1.079 &\quad 0.523 &\quad 0.523  &   \quad 1.270 & \quad 1.270 & \quad 0.786&\quad 0.206& M\\
	$P_9=(43, 10)$ & AIM & \quad 4.341 &\quad 0.325  &\quad 0.236 &\quad 0.696 &\quad 0.696  &   \quad 1.194 & \quad 1.194 &\quad 0.325 &\quad 0.236& AIM\\
	$P_{10}=(43, 0)$ & AIM & \quad 3.976 &\quad 0.391  &\quad 1.537 &\quad 0.807 &\quad 0.807 &   \quad 1.032 & \quad 1.032 & \quad 0.298&\quad 0.000& AIM\\
	\bottomrule
\end{tabular}
\end{table*}

Figs.~\ref{fig3}(b) and (c) present the numerical phase diagrams for a 2D and 3D mixture,
and Fig.~\ref{fig3}(a) shows the extended phase diagram for a quasi-1D mixture. All panels exhibit transitions between M and AIM phases as $\beta_{12}$ and $\beta_p$ vary. The M phase expands significantly with increasing the spatial dimension, which can be intuitively understood as
follows. For a quasi-1D BEC mixture in the M phase and near the phase boundary,
reducing the transverse confinement allows it to expand along the transverse direction, eventually forming a 2D mixture. Such transverse expansion further reduces the repulsive interspecies $s$-wave interaction, favoring the M phase. The 3D phase diagram resembles the 2D one, but with an even larger M phase region. For a specific $\beta_{12}$, the mixture undergoes a transition from the AIM to the M phase with increasing $\beta_p$. Therefore, (positive) $p$-wave interaction enhances the miscibility of the BEC mixture.

It is interesting to compare Fig.~\ref{fig3}(a) with Fig.~\ref{fig1}(a), both for the quasi-1D mixture. 
For small $\beta_{12}$ and $\beta_p$, Fig.~\ref{fig3}(a) aligns with Fig.~\ref{fig1}(a),
noting the differing vertical axis scaling in Fig.~\ref{fig3}. 
Notably, for $\beta_{12}$ surpassing a threshold ($\beta_{12}\geq\beta_{12}^0=6.0$ in 1D), the M phase disappears, and the ground state remains in the AIM phase regardless of $\beta_p$. Similar thresholds, $\beta_{12}^0$ values of 18.0 (2D) and 48.0 (3D), are observed [cf. Fig. 3(b) and (c)]. This phenomenon arises because neither the completely AIM phase $(\eta = 0, d > 0)$ nor the completely M phase $(\eta = 1.0, d = 0)$ receives energy contributions from $p$-wave interactions. Beyond the threshold of $\beta_{12}$, the energy reduction in the AIM phase due to $\beta_{12}$ surpasses the kinetic and potential energy reductions achieved by mixing. Therefore, the mixture enters the AIM phase consistently for sufficiently large $\beta_{12}$.
For a fixed value of $\beta_p$, we examine the depedence of the order parameters on $\beta_{12}$, as depicted in Fig.~\ref{fig4}.
As $\beta_p$ approaches zero (e.g., $\beta_p=0.1$),
the order parameters $\eta$ and $d$ evolve continuously with $\beta_{12}$ in 1D (blue solid line), 2D (red solid lines), and 3D (yellow solid lines).
However, at $\beta_p=1.0$, only in 3D do the order parameters vary smoothly with $\beta_{12}$ (yellow dashed line),
while in 1D and 2D, they exhibit discontinuous jumps. 
Increasing $\beta_p$ to 10.0 results in a similar discontinuous transition in 3D (purple dotted line). 
Thus, in all spatial dimensions, the phase transition is continuous for weak $p$-wave interaction and becomes first order for strong $p$-wave interaction. 

This transition results from the presence of metastable states, as extensively explored via the 1D Gaussian variational analysis in Sec.~\ref{sec:VariationalModel}. In the presence of $\beta_p$, the system features two metastable states with low symmetry (AIM phase). Their energy is associated with $\beta_p$, with larger $\beta_p$ corresponding to higher energy. In the M phase, the energy of these metastable states is higher than the completely mixed state. As $\beta_{12}$ increases, the energy of these states decreases, eventually becoming lower than that of the higher-symmetry mixed state. This results in a phase transition accompanied by a sudden change in the order parameter and a phase transition. Larger $\beta_p$ requires a higher $\beta_{12}$ for this transition due to the increased energy of metastable states.

Interestingly, the phase boundary in 3D becomes more intricate at the transition point between continuity and discontinuity. 
This complexity is evident in the inset of Fig.~\ref{fig3}(c), 
revealing a distinct peninsula pattern around the transition point due to the dual effect of $p$-wave interactions.
In contrast, while present in the 1D and 2D cases, this feature is less pronounced.
The energy changes at points $P_5$-$P_{10}$ [marked in Fig.~\ref{fig3}(c)] in Table~\ref{Table} reveal that, 
for $\beta_{12}=43$, variations in $\beta_p$ induce changes in the individual component energies of the binary BEC mixture. 
Regardless of $\beta_p$, the energy contribution from $p$-wave interactions consistently remains zero in the completely M phase, 
maintaining the total energy constant ($\varepsilon_{\rm tot}=4.552$ in Table~\ref{Table}). 
Thus, if the energy of the AIM phase is lower than this value, the true ground state should be in the AIM phase. 
Although the AIM configuration reduces the energy associated with $\beta_{12}$, it simultaneously increases the kinetic and potential energies of both BEC components, as well as the $p$-wave interaction energy introduced by incomplete separation of two BECs.

At $P_{10}$ ($\beta_p \gtrsim 0$) with a sufficiently large $\beta_{12}$ ($> \beta^0_{12}$), the ground state resides in the AIM phase, significantly reducing the energy introduced by $\beta_{12}$. As $\beta_p$ increases to $10^3$ at $P_5$, a transition to the completely M phase becomes accessible, nullifying dominant $p$-wave interaction energy and decreasing kinetic energy compared to the AIM configuration. However, this transition raises the $s$-wave interspecies interaction energy $\varepsilon_{\beta_{12}}$ from 0.001 to 1.164 due to wave function overlap. From $P_{10}$ to $P_8$, where $\beta_p$ increases from 0 to 20, the ground state undergoes a transition from the AIM to the M phase, with the order parameter $\eta$ rising from 0.391 to 1.00. Here, $p$-wave interaction lowers $\varepsilon_{\rm tot}$ and enhances mixing. Reaching $\beta_p = 27$ at $P_7$, the mixture reverts to the AIM phase to reduce the total energy $\varepsilon_{\rm tot}$, with a reduced $\eta=0.356$ and $d=1.917$. In this case, $p$-wave interaction enhances phase separation. We emphasize that the energy difference of the ground and metastable states is very small at points $P_6$-$P_9$ in Table \ref{Table}, and such intricacy results in a more complex phase boundary.

\begin{figure*}[t!]
	% \flushleft
	\centering
	\includegraphics[width= 2.0\columnwidth]{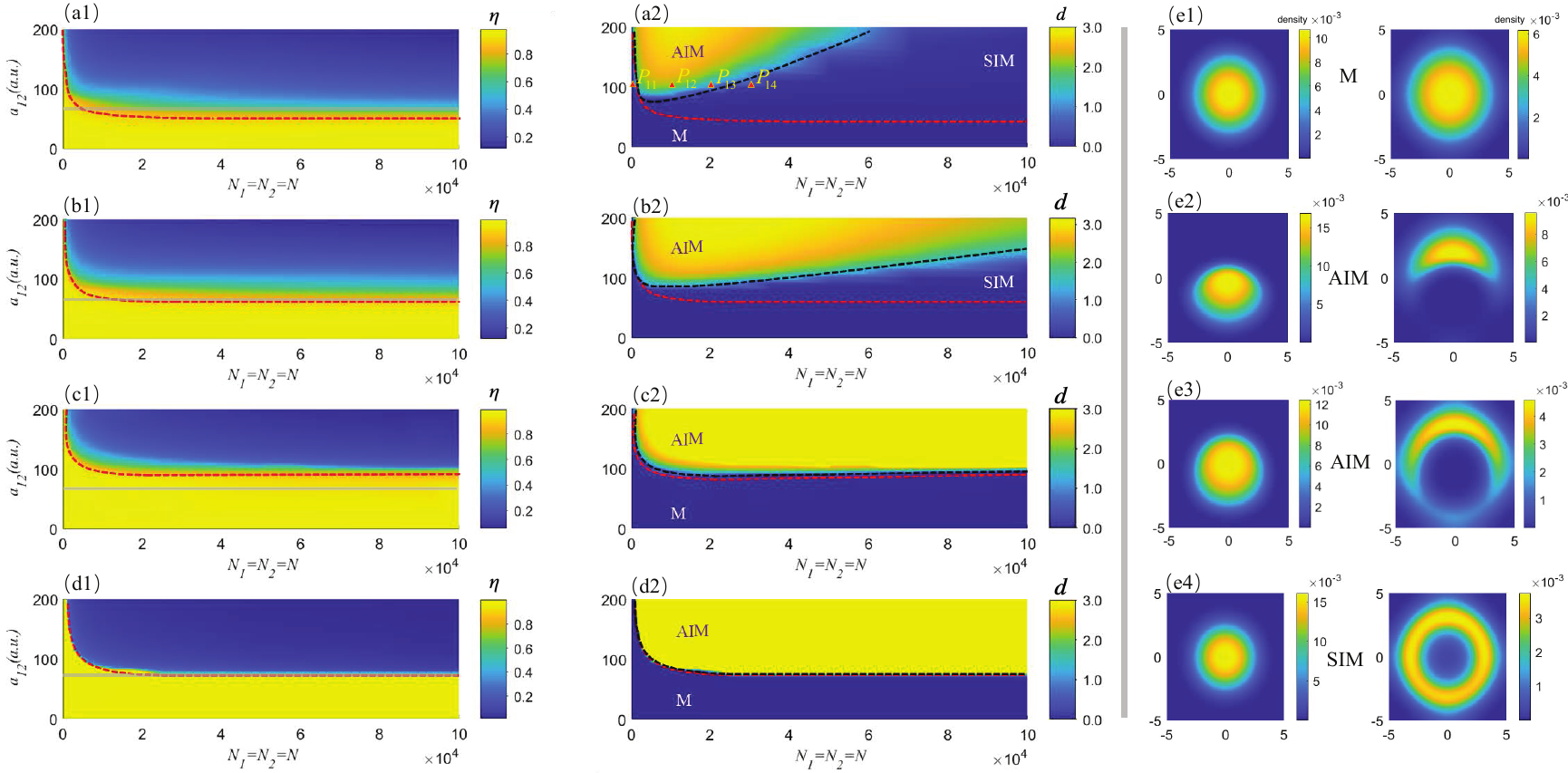}
	\caption{Ground-state phase diagram of a mixture of \Rb{87}-\Na{23} mixture in the $N$-$a_{12}$ plane. 
	The interspecies scattering lengths are $a_{11}=113.4~\mathrm{a.u.}$ and $a_{22}=57.0~\mathrm{a.u.}$.
	The $p$-wave interspecies interactions $\beta_p/\beta^{bg}_{12}$ is $0$ in (a), 0.2 in (b), 0.8 in (c) and 3.6 in (d), 
	the background scattering length $a_{12}^{bg} = 67~\mathrm{a.u.}$ (a1)-(d1) show the distribution of the order parameter $\eta$,
	while (a2)-(d2) show the distribution the order parameter $d$. The red and black dash lines represent the phase boundaries determined by numerical calculation, the light gray solid lines in (a1)-(d1) represents the critical value of $a_{12}=80~\mathrm{a.u.}$
	The density profiles of the two BEC components at locations $P_{11}=(5\times10^2, 100~\mathrm{a.u.})$, $P_{12}=(1\times10^4, 100~\mathrm{a.u.})$, $P_{13}=(1.8\times10^4, 100~\mathrm{a.u.})$ and  $P_{14}=(3\times10^4, 100~\mathrm{a.u.})$ in panel (a2) are shown as (e1)-(e4) respectively.}%
	\label{fig5}%
\end{figure*}

Notably, the SIM phase, frequently reported in the literature~\cite{wen21pra,Esry1999Spontaneous}, is absent in Fig.~\ref{fig3}. This is due to our assumption that both BECs share the same intraspecies interaction strength, have the same spherically symmetric harmonic trap and an equal number of atoms. Generally speaking, the interplay between M, SIM, and AIM phases primarily depends on the competition between interspecies $s$-wave interactions
and factors such as kinetic energy, potential energy, and intraspecies interactions---the former determines the miscibility, while the latter dictates the spherical symmetry of the ground state. Attaining an SIM phase involves investigating factors leading to distinct intraspecies interactions, including variations in masses, harmonic trap frequencies, diverse intraspecies $s$-wave scattering lengths, and the relative atomic number ratio of the two species, as detailed in the subsequent discussion.

\subsection{The ground-state phase diagram with intraspecies interactions} \label{sec:IntraspeciesModel}
Until now, our focus has been on a simplified model, neglecting intraspecies interactions ($\beta_1$ and $\beta_2$) and assuming equal masses and trapping frequencies for both species. In practical experiments, however, intraspecies interactions play a crucial role. Repulsive intraspecies interactions ($\propto\beta_i|{\phi_i}|^4$) tend to expand individual condensates, thus favoring the miscible phase~\cite{wen21pra,Lee2016Phase,bisset2018enhanced}. This effect is evident in Fig.~\ref{fig1}, where the nonzero intraspecies interaction $\beta_1=\beta_2=2$ significantly extends the boundary of the M phase at $\beta_1=\beta_2=0$.
Intraspecies interactions not only broaden density profiles for both species
but also trigger the emergence of a SIM phase ($d=0$, $\eta<1$)~\cite{wen21pra,Esry1999Spontaneous},
where the less-bound BEC encapsulates the other, resembling a symmetric spherical shell with lower energy and higher symmetry than the AIM phase.

Atom number plays a crucial role in determining the ground state symmetry and phase boundary structure of binary BEC mixtures,
as demonstrated in works that consider the only $s$-wave interaction~\cite{Esry1999Spontaneous,McCarron11Dualspecies,wen21pra,app11199099}.
The critical scattering length ($a^c_{12}$) for phase separation is found to be atom number-dependent,
as shown in the investigation of spontaneous symmetry breaking (SSB) in a trapped Na-Rb BEC mixture~\cite{Esry1999Spontaneous}.
Further investigation in Ref.~\cite{wen21pra} explores how atom number ratio ($\gamma=N_1/N_2$) influence miscibility-immiscibility transitions.
In a homonuclear BEC mixture involving two magnetic sublevels $m_F=\pm1$ of the hyperfine spin $F=1$ state,
both species share identical intraspecies $s$-wave scattering lengths, as demonstrated in the \Na{23}-\Na{23} BEC mixture~\cite{Fava2018Observation}.
Phase diagrams in the $\gamma - a_{12}$ plane display three distinct phases: M, AIM, and the SIM.
The absence of the SIM phase at $\gamma = 1$ aligns with our earlier numerical findings.
However, in a binary BEC mixture with differing intraspecies $s$-wave scattering lengths,
an SIM phase can emerge between the M and AIM phases at $\gamma=1$.
Below, we extend these discussion to the scenario where the interspecies $p$-wave interaction is present.

\begin{figure*}[t!]
	% \flushleft
	\centering
	\includegraphics[width= 2.0\columnwidth]{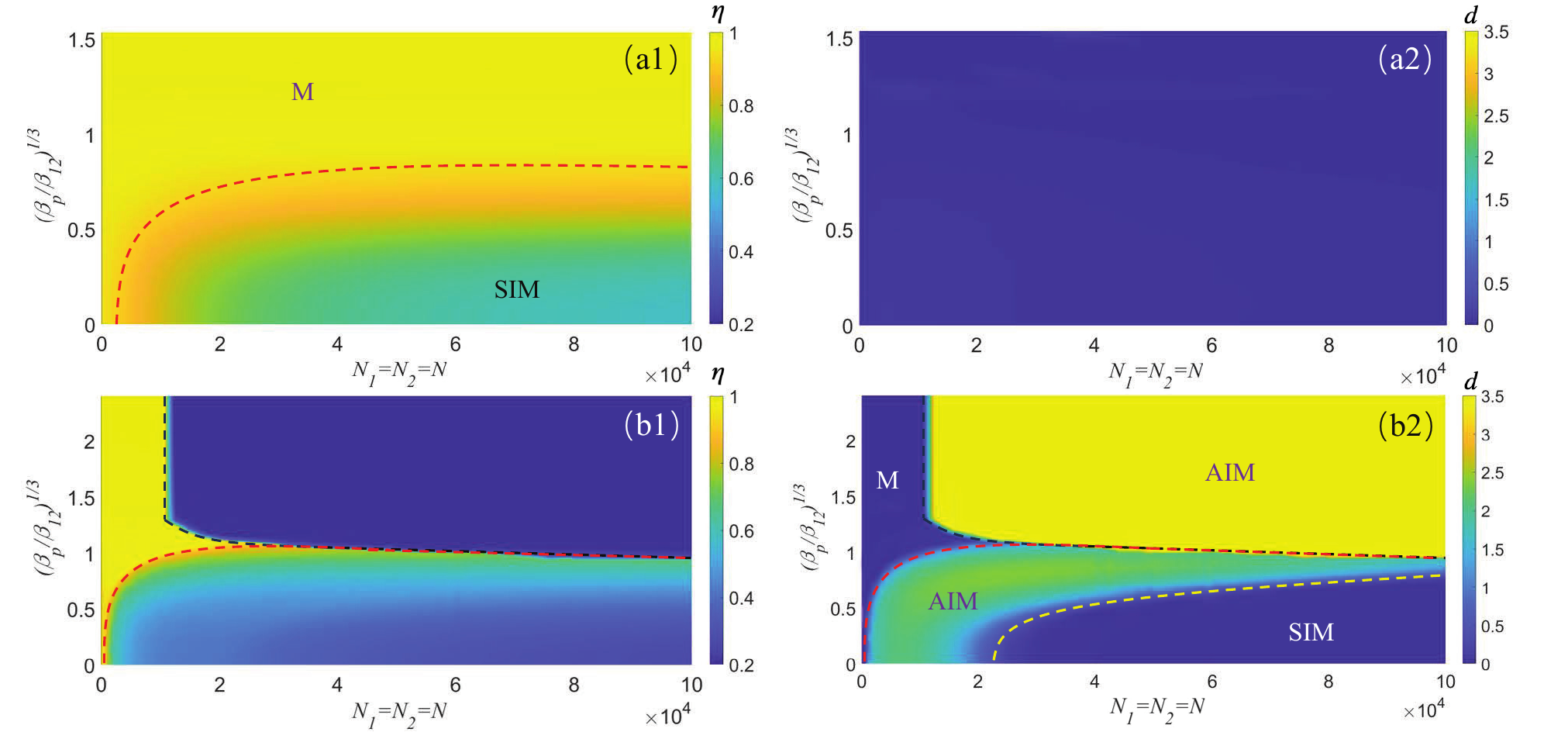}
	\caption{Ground-state phase diagram for the \Rb{87}-\Na{23} mixture
	with fixed $a_{11}=113.4~\mathrm{a.u.}$ and $a_{22}=57.0~\mathrm{a.u.}$ in the $N$-$(\beta_p/\beta^{bg}_{12})^{1/3}$ plane.
	The value of $s$-wave interspecies interaction in (a) is the background scattering length $a_{12}=67~\mathrm{a.u.}$,
	while in (b), it is set to $a_{12}=100~\mathrm{a.u.}$. (a1) and (b1) correspond to the order parameter $\eta$, while (a2) and (b2) correspond to the order parameter $d$. The red, black and yellow dashed lines represent the phase boundaries given by numerical imaginary-time evolution of the GP equation.}%
	\label{fig6}%
\end{figure*}

Our numerical simulation employs parameters close to a broad $p$-wave FR at approximately $284$\,G
in the $^{87}\text{Rb}\,|{1, + 1}\rangle + ^{23}\text{Na}\,|{1, + 1} \rangle $ channel~\cite{wang2013observation}. Two $s$-wave resonances at 347.8 and 478.8G have been also observed
near this $p$-wave resonance.
% Assigning \Rb{87} as species 1 and \Na{23} as species 2, we adopt an isotropic harmonic trap,
% setting $\omega_1 =(2\pi) 160$ Hz for the Rb atoms and $\omega_2 = \omega_1\sqrt{m_1/m_2}\approx 2\omega_1$
% for Na atoms to ensure equal effective trap potentials in Eq.~(\ref{eq:eqEnergy3}),
% as previous studies~\cite{Esry1999Spontaneous}.
Hereafter, we only consider the case of $\gamma=1$ and take the $s$-wave scattering length $a_{11} = 113.4~{\rm a.u.}$
and $a_{22}=57.0~{\rm a.u.}$ as constants independent of the magnetic field~\cite{volz2003characterization,knoop2011feshbach}.
By minimizing the energy functional of Eq.~(\ref{eq:eqEnergy2}) for the 3D case, we arrive at the phase diagram presented in Fig.~\ref{fig5}.
From Fig.~\ref{fig5}(a) to Fig.~\ref{fig5}(d), the $p$-wave interaction strength $\beta_p/\beta^{bg}_{12}$ is chosen increasingly, as $0.0, 0.2, 0.8$ and 3.6; the background $s$-wave scattering length is fixed at $a_{12}^{bg} = 67~\mathrm{a.u.}$~\cite{wang2013observation}.

Without the $p$-wave interaction [Fig.~\ref{fig5}(a)],
the critical interspecies scattering length is $a^{c0}_{12}=62.5~\mathrm{a.u.}$, beyond which phase separation occurs in the TFA.
However, results from Fig.~\ref{fig5}(a1-d1) indicate a significant influence to $a^c_{12}$ by of the atom number $N$.
For small $N$, $a^c_{12}$ roughly scales with $N^{-1}$, whereas for large $N$,
the $N$ dependence of $a^c_{12}$ remains nearly constant due to the substantial differences in intraspecies interactions.
These distinctions give rise to a shell-filling effect for the Na atoms~\cite{Esry1999Spontaneous},
defining the shell as the spatial region between the mean ield of the more tightly trapped Rb atoms and the less confined Na atoms,
manifesting as the SIM region in Fig.~\ref{fig5}(a2) and (b2).
However, if $a_{12}$ falls below the critical value of $a^c_{12} \approx 80~\mathrm{a.u.}$ [cf. the light gray solid lines in Figs.~\ref{fig5}(a1)-(d1)],
the mixture remains in the M phase due to the dominant intraspecies repulsive interactions.
Upon surpassing this threshold, the mixture undergoes a configuration transition towards the SIM phase,
as interspecies interactions and other terms begin to dominate over intraspecies interactions.
Specifically, the Rb atoms at the trap center gives rise to a spherically symmetrical potential barrier to the Na atoms; the latter thus forms a symmetrical shell structure.
Upon increasing $a_{12}$, the wave function overlap ($\eta$) decreases.
Beyond the phase boundary (black dashed line) in Fig.~\ref{fig5}(a2),
interspecies interaction becomes dominant.
To minimize interaction energy, the two species disperse in opposite directions due to the strong repulsive interaction.
The symmetry-broken ground state features an approximately planar boundary between the two hemispheres of each species [Fig.~\ref{fig5}(e2)],
reducing both kinetic energy from boundary curvature and interface volume.
In this parameter regime, the system undergoes SSB, entering the AIM phase~\cite{Esry1999Spontaneous}.

Furthermore, we focus on the case of $a_{12}=100~\mathrm{a.u.}$ in the absence of $p$-wave interaction.
The variation of the mixture's ground-state phase with increasing $N$, labeled as $P_{11}$-$P_{14}$ in Fig.~\ref{fig5}(a2),
is depicted in Figs.~\ref{fig5}(e1)-(e4).
At small $N$, the dominance of kinetic energy forms a significant barrier hindering phase separation.
Consequently, the system initially adopts a symmetric M phase [$P_{11}$ with Fig.~\ref{fig5}(e1)].
As $N$ increases, interaction energy rises faster than kinetic energy,
as a result of the higher order scaling of interaction energies with respect to $N$ than kinetic energy.
The energy loss associated with phase separation, driven by the interspecies interactions between two species,
compensates for the gain in kinetic energy, favoring a AIM phase [$P_{12}$ with Fig.~\ref{fig5}(e2)].
As $N$ continues to increase (while still remaining in the AIM phase), the Na atoms wrap further around the Rb core.
The Rb core is pushed slightly off the trap center, but the trap is so tight such that large displacements are energetically unfavorable [$P_{13}$ with Fig.~\ref{fig5}(e3)]. Eventually, as the number of atoms grows sufficiently, the Na atoms completely encompass the Rb core by filling the available shell-like space.
Thus, the mixture enters the SIM phase  [$P_{14}$ with Fig.~\ref{fig5}(e4)].

The $p$-wave interspecies interaction has dual effects:
Figs.~\ref{fig5}(a1-c1) show that $a^c_{12}$ increases to about $100~\mathrm{a.u.}$ as $p$-wave interaction increases to $\beta_p=0.8\beta^{bg}_{12}$.
In this regime, the $p$-wave interaction enhances mixing of the two BECs.
However, when $\beta_p$ further increases to $3.6\beta^{bg}_{12}$ in Figs.~\ref{fig5}(d1) and (d2), $a^c_{12}$ decreases. In this regime, $p$-wave interaction enhances phase separation.
This is similar to the phenomenon in Fig.~\ref{fig3} (c).
Another consequence of $p$-wave interaction is that the SIM phase region gradually decreases as the $\beta_p$ increases,
as shown by the changes in the order parameter $d$ in Figs.~\ref{fig5}(a2) -(d2).
Considering a certain position in the SIM phase near the AIM phase boundary in Fig.~\ref{fig5}(b1), such as position $P_{14}$.
Although the spherical shell configuration reduces the kinetic energy and $s$-wave intraspecies interaction by reducing the boundary curvature,
it also increases the interface volume of the two-component BEC.
Simultaneously, the additional $p$-wave interaction tends to the reduction of the interface volume,
resulting in a rightward shift of the AIM phase boundary.
As $\beta_p$ increase the AIM phase expands until the SIM phase region is no longer visible in Figs.~\ref{fig5}(c2)-(d2).
In addition, the order parameters become discontinuous as $\beta_p$ becomes sufficiently large in Figs.~\ref{fig5}(d1) -(d2), the underlying mechanism has been  explained in Sec.~\ref{sec:VariationalModel}.

Figure~\ref{fig6} compares the ground-state phase diagram in the $(\beta_p/\beta^{bg}_{12})^{1/3}-N$ plane at different $a_{12}^{bg}$:
$a_{12}^{bg}=67\,\mathrm{a.u.}$ and $a_{12}^{bg} = 100~\mathrm{a.u.}$
Significant differences emerge when using $p$-wave interactions to modulate the miscible-immiscible phase transition.
In Figs.~\ref{fig6}(a), where $a_{12}^{bg} < a^c_{12}$, the system exclusively exhibits M and SIM phases.
The addition of $p$-wave interaction mainly facilitates the mixing of the two BECs, driving a transition from the SIM phase towards the M phase.
As discussed above, the  AIM phase exists only if $a_{12}^{bg}$ is sufficiently large.
Conversely, when $a_{12}^{bg}\ge a^c_{12}$, as observed in Figs.~\ref{fig6}(b), all three phases exist.
When $N \lesssim 10^4$, interspecies interaction ($\propto\beta_{12}$) is weaker than kinetic energy, the latter promotes mixing.
The inclusion of $p$-wave interaction enhances mixing, driving the AIM phase at $\beta_p = 0$ to the M phase.
As $N$ increases, the M-to-AIM phase boundary approximates a straight line, akin to the steep boundary observed when $\beta_{12}$ surpasses the critical value $\beta^c_{12}$ in Fig.~\ref{fig3} .
However, when $N > 10^4$, the $p$-wave interaction manifests dual effects:
weak $\beta_p$ promotes mixing and strong enough $\beta_p$ triggers phase separation.
With increasing $\beta_p$, the order parameter $\eta$ first increases and then decreases, as clearly depicted in Fig.~\ref{fig6}(b1).
Upon reaching $N > 10^5$, the mixture enters the TFA regime ($\beta_{1(2)} \gg 1$).
The ground state undergoes SSB around $\beta_p/\beta^{bg}_{12} \approx 1.0$,
entering directly the AIM phase. Correspondingly, the order parameter $d$ exhibits an
abrupt change, increasing from 0 to $d\ge3$~\cite{trippenbach2000structure}. 

Finally, we note that while we have primarily focused on investigating the role of $p$-wave interaction to the miscible-immiscible phase transition, other factors such as particle number ratio and trap configuration may also slightly influence the phase boundary~\cite{wen2012controlling,app11199099,wen21pra}. The detailed analysis of these is beyond the scope of this study.

\section{Conclusions and Discussions}\label{sec:conclusion}
In summary, we have derived a mean-field equation incorporating $p$-wave interactions by employing a 
single-channel $p$-wave pseudopotential model. 
Our exploration of the ground-state phase diagram in a binary Bose-Einstein condensate (BEC) with 
both interspecies $s$-wave and $p$-wave interactions has revealed richer characteristics than those arising 
from $s$-wave interactions alone. 
Positive $p$-wave interactions exhibit a dual impact, 
either promoting the mixing or driving the separation of BEC components. 
This introduces a novel avenue for experimentally controlling the mixing behavior of BEC mixtures, 
suggesting potential directions for future studies. 
In practical \Rb{87}-\Na{23} BEC mixture experiments, 
achieving precise control over both $s$- and $p$-wave interactions is challenging. 
However, specific magnetic field positions present a favorable scenario where a 
particular $s$-wave Feshbach resonance (FR) and a specific $p$-wave FR are closely aligned 
in diverse binary mixtures~\cite{Dong2016Observation,Cui2018Broad}.

In the broader scope of diverse mixtures involving various atomic species, 
multiple factors such as the atom number ratio, mass imbalance, 
and disparities in trapping configurations among the constituents could potentially influence miscibility\cite{wen21pra,app11199099}, 
providing avenues for future investigation. 
Additionally, it's important to highlight that our focus has primarily centered on the ground-state 
properties of BEC mixtures. The dynamics, especially the evolution of phase separation, 
remain unexplored in our present research~\cite{hall1998dynamics,mertes2007nonequilibrium,wirthwein2022collision,cavicchioli2022dipole}. 
Given that $p$-wave interactions involve gradient correlations between wave functions of the two components, 
investigating the phase separation dynamics prompted by $p$-wave interactions holds promise for future investigation.

\section*{Acknowledgments}
We would like to thank Prof. Li You for invaluable guidance and discussion.
% \nocite{*}

\bibliography{spwave}% Produces the bibliography via BibTeX.

\end{document}